\documentclass[aps,prl,twocolumn,showpacs,superscriptaddress]{revtex4-1}

\usepackage{graphicx}
\usepackage{color}
\usepackage{bm}
\usepackage{footnote}
\usepackage[bottom]{footmisc}

\begin{document}
\title{Giant isolated attosecond pulses from two-color laser plasma interactions}

\author{Y. X. Zhang}
\affiliation{Center for Applied Physics and Technology, HEDPS, SKLNPT, and School of Physics, Peking University, Beijing, 100871, China}
\affiliation{Helmholtz Institute Jena, Jena 07743, Germany}
\affiliation{Laser Fusion Research Center, China Academy of Engineering Physics, Mianyang, 621900, China}
\author{S. Rykovanov}
\affiliation{Helmholtz Institute Jena, Jena 07743, Germany}
\author{Mingyuan Shi}
\affiliation{Helmholtz Institute Jena, Jena 07743, Germany}
\affiliation{Institute of Optics and Quantum Electronics, Friedrich Schiller University， Jena 07743， Germany}
\author{C. L. Zhong}
\affiliation{Center for Applied Physics and Technology, HEDPS, SKLNPT, and School of Physics, Peking University, Beijing, 100871, China}
\author{X. T. He}
\affiliation{Center for Applied Physics and Technology, HEDPS, SKLNPT, and School of Physics, Peking University, Beijing, 100871, China}
\affiliation{Collaborative Innovation Center of IFSA, Shanghai Jiao Tong Univ., Shanghai 200240, China}
\affiliation{Institute of Applied Physics and Computational Mathematics, Beijing 100094, China}
\author{B. Qiao}
\email[Correspondence to: ] {b.qiao@pku.edu.cn}
\affiliation{Center for Applied Physics and Technology, HEDPS, SKLNPT, and School of Physics, Peking University, Beijing, 100871, China}
\affiliation{Collaborative Innovation Center of IFSA, Shanghai Jiao Tong Univ., Shanghai 200240, China}
\author{M. Zepf}
\email[Correspondence to: ] {m.zepf@uni-jena.de}
\affiliation{Helmholtz Institute Jena, Jena 07743, Germany}
\affiliation{Institute of Optics and Quantum Electronics, Friedrich Schiller University， Jena 07743， Germany}

\begin{abstract}
	
	A new regime in the interaction of a two-colour ($\omega$,$2\omega$) laser with a nanometre-scale foil is identified, resulting in the emission of extremely intense, isolated attosecond pulses - even in the case of multi-cycle lasers.  For foils irradiated by lasers exceeding the blow-out field strength (i.e. capable of fully separating electrons from the ion background), the addition of a second harmonic field results in the stabilization of the foil up to the blow-out intensity. This is then followed by a sharp transition to transparency that essentially occurs in a single optical cycle. During the transition cycle, a dense, nanometre-scale electron bunch is accelerated to relativistic velocities and emits a single, strong attosecond pulse with a peak intensity approaching that of the laser field.

\end{abstract}

\maketitle
\renewcommand\thesection{\arabic{section}}

Attosecond pulses are typically generated using phase locked frequency combs  (high-order harmonics or HHG) through non-linear processes in atomic gases. Attosecond pulses are widely used in many fields \cite{application, Pertot:2017,  Atta:2017, Hassner,  Dobosz, Lara-Astiaso:2018, Calegari:2014, Calegari:2014, Takahashi:2015, Review, Review1, Review2}, particularly in ultrafast atomic and molecular dynamics \cite{Pertot:2017,  Atta:2017}, characterizing plasmas by time-resolved X-ray diffraction\cite{Hassner,  Dobosz} or extreme ultraviolet (XUV)/X-ray pump-probe techniques \cite{Lara-Astiaso:2018,  Calegari:2014, Takahashi:2015}. For all these applications, the intensity and the degree of isolation of the single attosecond pulses are of great importance \cite{Review, Review1, Review2}. They are widely used to achieve attosecond temporal resolution  by employing a femtosecond excitation pulse to trigger the desired phenomenon followed by an attosecond pulse to probe the evolution of the system under study (femto-pump/atto-probe). High-intensity isolated attosecond pulses would allow higher temporal precision and a wider range of phenomena to be studied through atto-pump/atto-probe experiments.  The extension of the pump-probe techniques to the extreme ultraviolet (XUV) and soft x-ray regime based on the availability of intense attosecond pulses would open the way to exciting new territory such as real-time observation of a wide range of phenomena involving fast electron dynamics (including inner-shell dynamics). Other examples are  freeze-framing correlated electron dynamics or vibrational states\cite{Isolated application, Calegari:2014, Takahashi:2015}. \\
Using state-of-the-art multi-terawatt and petawatt class laser systems to increase the photon flux of attosecond sources seems an obvious choice, however HHG in atomic gases cannot efficiently exploit such lasers as they optimise at relatively low intensity \cite{Tsakiris}. Using a plasma-vacuum interface created on the surface of a solid target allows an intense XUV pulse to be created through a number of mechanisms known as Coherent Wake Emission (CWE\cite{Quere}), Relativistically Oscillating Mirror (ROM \cite{Lichters:1996, ROM Baeva:2006})  and Coherent Synchrotron Emission (CSE  \cite{CSE Pukhov, CSE Dromey}). In the following we focus on the intense CSE radiation emitted from thin foil plasmas.  
  If  CSE can be restricted to a single event, e.g. by using a  single cycle pulse, the emission of a strong isolated attosecond pulse occurs \cite{Cousens}. In general, however, multi-cycle drive lasers result in the emission of an attosecond pulse every (half-) laser cycle and thus an attosecond pulse-train \cite{AS train}. Thus for multi-cycle interactions  the non-linear process must generally be gated with a suitable approach to achieve a single, dominant attosecond pulse \cite{Review, Review1, Review2, Silva:2015, Intensity gating, Polarisation gating, Two color gas}. This is true for both HHG in gases or plasma surfaces \cite{CSE Dromey, Nomura:2009, Lichters:1996, Gibbon:1996, ROM Baeva:2006, ROM Dromey}.\\

The temptation is to find a route to converting lasers at the cutting-edge of achievable peak power (Petawatt-scale) into isolated attosecond pulses. The interaction of a laser with highly relativistic intensity ($a_0=I\lambda^2/1.3 \times 10^{18} \rm{W}/\rm{cm}^2 \gg1$) with a plasma surface has been identified as a potential route to isolated, high-energy attosecond pulses \cite{Vincent:2012, Wheeler:2012, CEP, Zhang:2018, Yeung:2014, Sergey:2008} and developing new laser technologies with sufficiently short PW pulses is the subject of intense efforts \cite{Laser system}.
To date, lasers with such high peak power are multi-cycle lasers with pulse durations exceeding 20fs. Similar gating techniques to those mentioned above are applicable for relativistic plasmas and short pulses ($\tau \approx10\rm{fs}$) \cite{Vincent:2012, Wheeler:2012, CEP, Zhang:2018}.  Polarization gating has been proposed for longer laser pulses ($\tau>10\rm{fs}$) \cite{Yeung:2014, Sergey:2008}, but none of these approaches have allowed a single isolated attosecond pulse to be demonstrated yet. \\
In this letter, we highlight an entirely new regime of laser-plasma interaction suitable for converting multi-cycle lasers to a single attosecond pulse. Two colour interactions \cite{Edwards:2014, Yeung:2017} with the correct relative phase of $\pi$ lead to an essentially single cycle transition in the target state from highly reflective to relativistically transparent\cite{transparency}. This unique dynamic implies that there is a sharp transition over one cycle with a very large proportion of the foil electrons emitting coherently during the transition cycle, resulting in a giant attosecond pulse with a peak intensity approaching that of the driving laser.
 
 The  paper is structured as follows: First we consider the difference between one and two-colour interactions, highlighting the fact that in two colour interactions the foil remains dense and stable up to a single cycle where the foil becomes transparent. We then focus on the dynamics that make the `blow-out' cycle (at which the foil becomes transparent) unique in terms of emission strength. Finally, we give discussions on the optimal parameters and the robutness of this approach on the single attosecond pulse generation.

 \section{Difference between one and two-colour interactions}
   
CSE is well known from accelerator rings and occurs when an electron bunch is confined to a spatial scale of less than one wavelength in the direction of motion and accelerated. The situation when a laser interacts with electrons in a foil is analogous, with the acceleration taking place in a single laser cycle leading to bright burst of radiation \cite{transparency2, Bulanov2013}. For a bunch with spatial extent $<\lambda/2$ in the radiation of each electron at wavelength $\lambda$ adds constructively at the observer and hence resulting in intense synchrotron radiation. In the XUV this requires a dense electron bunch confined on a nanometer-scale. Producing a single, bright pulse requires a high-charge bunch to be accelerated strongly in one single laser cycle. For multi-cycle lasers, adjacent cycles have similar intensity and therefore generally have very similar bunching and acceleration, resulting in attosecond pulse trains.

In the following we describe the emission of a single, dominant attosecond pulse in two-colour interactions. As will be shown, two colour interactions significantly reduce electron loss from the foil. This combines with the unique characteristics of the blow-out cycle (when $ a > \pi (n_e / n_c) (d/\lambda) $  \cite{transparency}) where the laser is strong enough to overcome the electrostatic force of the ions and separate the electrons from the foil: A single, high charge nm-scale electron bunch is formed resulting in intense XUV emission.

\begin{figure}
\includegraphics[width=7.5cm]{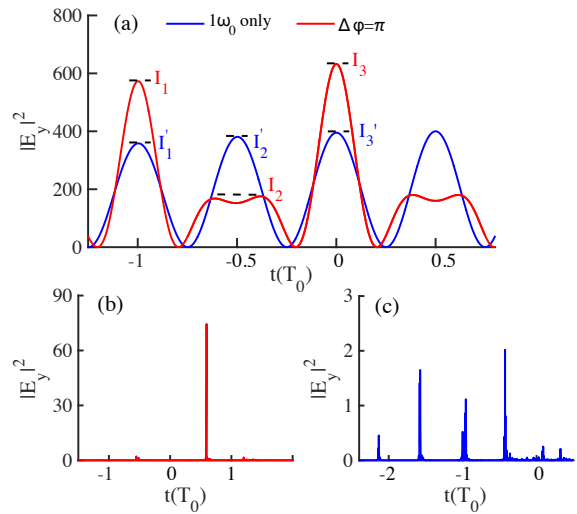}
\caption{(color online) 1D simulation highlighting the difference between two-colour with one-colour interactions. (a) shows the temporal intensity variation where $E_y$ is in units of $m_e\omega_0c/e$.   The reflected XUV emission (filtered from $30$-$200\omega_0$) in (b) and (c) shows a single attosecond burst for the TC case which is about $\times 40$ brighter than the OC  pulse-train.  }
 \label{fig:first}
\end{figure}

The distinction between the one-colour (OC) and two-colour (TC) cases  (for a relative phase of $\pi$) is clearly visible in 1D particle-in-cell simulations performed with PICWIG1D (resolution $\lambda/2000$ and 1000 quasiparticles per cell). Simulation parameters were linear polarization at normal incidence with Gaussian temporal profiles of $f(x) \sim \exp[-2\ln2(\frac{t-x-t_0}{\tau})^2]$ for the fundamental $E_{y,1}\sim a_0\sqrt{1-W}f(x)\sin(\omega t-kx)$ and the second harmonic $E_{y,2}\sim a_0\sqrt{W}f(x)\sin[2(\omega t-kx)+\Delta \phi]$ with $a_0=eE_y/m_e\omega_0 c=20$ and duration $\tau=20\rm{fs}$. For the TC case  $\Delta \phi =\pi$ and the energy ratio was $W=E_\omega / E_{2\omega}=0.1$. Carbon Foils with thickness $d = 8\rm{nm}$ initially locate at $x=0$ and $t=0$ is set as the moment the peak intensity reaches here. The foil density was adjusted to keep the blow-out parameter $\epsilon=\frac{\pi n_{e,0}d}{a_{max} n_c \lambda}$ constant (blow out at the peak laser cycle) \cite{transparency2}: $n_{e,0} = 600n_c$ for TC and  and $n_{e,0}=480n_c$  for OC.

 The primary difference is in the waveform interacting with the foil. As shown in figure \ref{fig:first}, the two color case exhibits alternating intense and weak ($I_1, I_2$) peaks instead of every half-cycle reaching the same intensity. Another point of difference is that the strong cycle rises and falls slightly faster, while the weak cycle exhibits $\approx 0.4$ cycles of near constant intensity.  The effect on the temporal emission pattern and intensity is dramatic: instead of many  attosecond XUV bursts with comparable intensity a single burst with 40x greater intensity.  

\begin{figure}
\includegraphics[width=8.5cm]{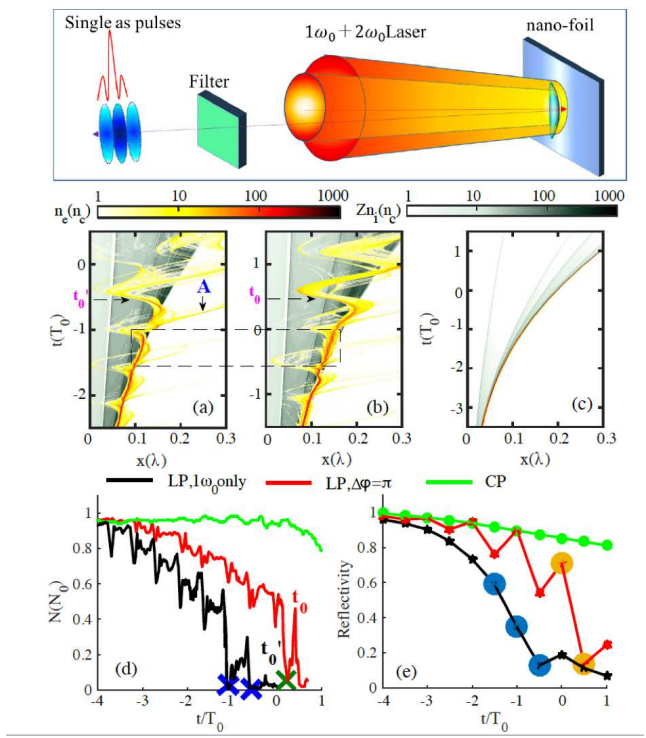}
\caption{(color online) Thin foil dynamics and foil blow-out process. Panels (a)-(c)  show ion and electron density for OC, TC and CP cases respectively. The  corresponding number of foil electrons (located $\pm d/2$ of the foil centre)  and laser reflectivity R is shown in (d) and (e).  Note the RPA -type behaviour (smooth acceleration of electrons and ions) in (a) while the electrons oscillate around the ion background (b) and (c). The dashed rectangle region highlights the different oscillation  dynamics prior to the blow-out cycles ($t_0^{'}$ in (b) and $t_0$ in (c)). The strong oscillations correspond to the maximum emission moments at the foil blow-out half-cycle for OC and TC cases, respectively. The OC case exhibits rapid electron loss before the blow out cycle (highlighted  by  $A$ ). At the blow-out cycle at which strongest CSE occurs only $0.14N_0$ are available  in the OC case in contrast to $0.45N_0$ for TC.  This higher electron density is the result of the reduced electron loss and rebunching  during the quasi-static low intensity TC half-cycles. Rebunching is clearly visible in the recovering reflectivity for TC in (e). Further evidence of the higher electron density in the TC case before blow out is the  steep single cycle drop from $R=0.7$ to $R=0.1$ during the blow-out cycle. }
 
 \label{fig:second}
\end{figure}

To explain the significantly different results obtained in TC case, we consider density evolution and reflectivity dynamics of the laser irradiated thin foil for different cases. Figure \ref{fig:second}(a)-(c) shows the spatio-temporal evolution of  electron and ion densities ($n_e$ and $n_i$) for  linearly polarized OC  and TC  cases and  circularly-polarized (CP) light, respectively.  For the linearly polarized OC case the Lorentz force $f_L\sim v\times B$ of the laser acting on electrons oscillates with a periodicity of half a cycle: Electrons are pushed from their equilibrium positions into the plasma when $f_L$ increases, while the Coulomb force $f_C$ caused by the displacement of the electrons pulls them back when $f_L$ decreases, resulting in a relativistic electron bunch oscillating around the ion background. It is this bunch that emits in attosecond pulse train in figure  \ref{fig:first}(c). Note that the highest energy electrons in each cycle do not return to the foil as highlighted by the box marked `A' in figure \ref{fig:second}(a) and a fraction of the relativistic bunch leaks out of the foil in a manner familiar from plasma surfaces \cite{Lichters:1996, Gibbon:1996, ROM Baeva:2006, ROM Dromey}.  The electron loss results in a falling laser reflectivity clearly visible in (d) and (e), with a sudden transition to transparency with the remaining foil electrons being detached at the blow-out cycle. After the blow-out cycle no well defined nm-scale bunch forms and further XUV emission is negligible. Note that despite the low number of electrons remaining in the OC blow-out cycle, the strong acceleration results in a final CSE emission burst slightly stronger than previous cycles. The trajectory of the blow out cycle is unique and leads to large acceleration of the remaining electrons in a narrow bunch.

For the TC case  at first glance the situation looks similar. However, significant differences have already been highlighted by some researchers investigating thick solid targets. Denser bunches and stronger acceleration  have been reported when a two-colour laser pulse interacts with an overdense plasma \cite{Mirzanejad:2013, Edwards:2014, Yeung:2017},  showing that two-colour dynamics differ greatly from a single frequency interaction. The trend to higher acceleration and bunch density due to the changed wave-form is also observed in thin foils, enhancing the strength of the CSE emission.  In the case of thin foils a further, very significant difference arises. The TC rate of electron loss in  cycles preceding blow-out intensity is greatly mitigated in two-colour interactions (See the rectangle region in \ref{fig:second}(b)) and (d),(e). As seen figure   \ref{fig:first} the low intensity half-cycle in the TC case ($I_2$) exhibits a long time (~$0.4$ cycles) of only slowly varying intensity. This results in a quasi-static effective potential allowing the remaining electrons to re-bunch. The density of the electrons in the vicinity of the foil increases during the quasi-static phase as shown by the periodic reflectivity increase for the TC case in (e). No significant electron loss occurs during the weak half-cycles, halving the number  electron loss events  compared to the OC case.  This has a pronounced effect on the reflectivity where figure  \ref{fig:second}(e) shows the energy ratio of reflected and incident pulses in each half-cycle. In the OC case the reflectivity drops smoothly from $R=0.7$ to $R=0.1$  over 4 half cycles, in the TC case abruptly in a single half-cycle. Also note that $R$ recovers during the weak TC half-cycles, providing clear evidence of the rising electron density due to re-bunching in the quasi-static half-cycle. Therefore, compared to the OC case, the effect of these weak,  slowly varying TC half-cycles is to reduce the rate of electron loss, and to change the foil dynamics to a sudden well defined blow-out cycle followed by single-cycle transition  from high reflectivity to relativistic transparency \cite{transparency}.\\
The CP case highlights the difference between the two cases made by the quasi-static half-cycle, by providing a comparator where there is no oscillating force component normal to the target and hence almost no electron loss despite the strong field. For CP, $|E|$,$|B|$ follow the pulse envelope resulting in quasi-constant ponderomotive push from the Lorentz force  $f_L\sim {\bf v \times B}$. The  force balance with $f_C$ results in a narrow, compressed electron sheath,  smooth ion acceleration (c) and a  low rate of electron loss from the foil and near-constant reflectivity as shown in (d) and (e).This dynamic has been widely discussed in the context of radiation pressure acceleration of ions (RPA \cite{Robinson}). 

The blow-out cycle results in the most intense CSE emission  for TC and OC at time $t_0$ and $t_0^{'}$ respectively. However,  $0.45 N_0$ electrons remain in the TC compared to only about $0.14 N_0$ left in OC, where $N_0$ is the initial number of electrons, making a significant contribution to the stronger XUV emission observed with TC. The sharp blow-out transition also suppresses CSE for all subsequent cycles, as there is no longer a well-localised nm-thin electron sheet confined around the ion background.   This is clearly visible in Figure \ref{fig:second} (b,c): the red electron density colours no longer track the densest part of the ion distribution after cycle 58.5. Before the blow-out cycle  (a large proportion of)  the electrons can be seen to be confined to the foil ions, while the electrons are essentially oscillating freely in the laser field  in a large bunch after the cycle . The absence of a effective foil, and therefore bunching, clearly gives strong non-linear suppression of CSE emission after the  blow-out cycle. However, this does not explain why the blow-out cycle results in an isolated attosecond pulse in the TC case (and the strongest pulse in the OC case despite the electron loss). To understand why the blow-out cycle is unique in being ideal for strong CSE emission compared to all previous cycles we must look at the bunching dynamics.

\begin{figure}
\includegraphics[width=6.5cm]{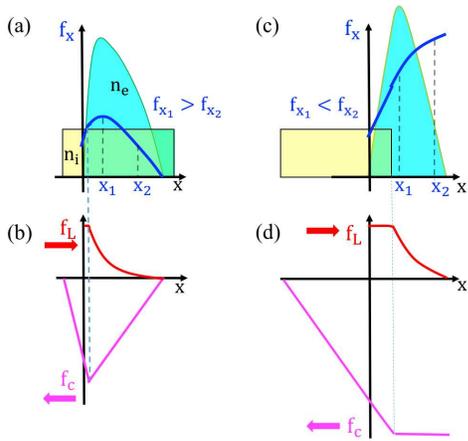}
\caption{(color online) Bunch compression before (a,c) and during the blow out cycle (b,d).  The forces applied to the electrons by laser and Coulomb forces before (a) and during  during the blow-out cycle (b) are shown in cartoon form to highlight the differences. The ions (yellow), electrons(blue) and resultant force $f_x$ are shown above (a,b), with the laser force  $f_L=-ev_yb_z$ and the Coulomb force $f_C$ below (c,d).}
 \label{fig:third}
\end{figure}

Figure \ref{fig:third} compares the different longitudinal force structures acting on the electrons in the pre blow-out and blow-out cycles. There are two major components to the force acting on the foil electrons, the force of the laser ($f_L$) normal to the target (mediated via the  $v \times B$- term of the Lorentz-force) and  the charge separation force $f_C$. The shape of the force that the electrons experience before and during the blow-out cycle  is significantly different. Before the blow-out cycle, the laser field decays exponentially ($\sim e^{-x/\sigma}$) in the electron layer over the skin depth ($\sigma$), leading to an exponentially-decaying force $f_L$. Thus, electrons at position $x=x_1$ feel a larger force $f_x=f_C-f_L$ than those deeper in the target at $x_2$. This results in a relative velocity $v_{x_1}>v_{x_2}$ after acceleration. Consequently, the electrons debunch from their initial high density dictated by the confining potential of the foil ions and the electron density $n_e$ drops as the electrons propagate.

\begin{figure}
	\includegraphics[width=8.5cm]{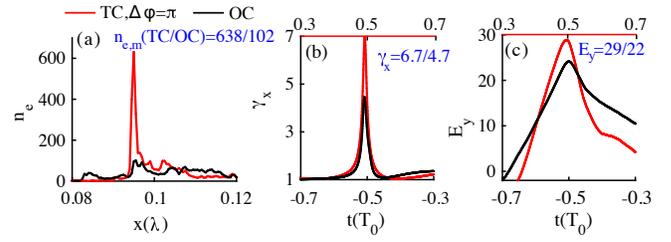}
	\caption{(color online) Bunch and field characteristics at the blow-out cycle. (a) electron density $n_e$ and (b) $\gamma_x$ and $E_y$ (units of $e/m_e\omega_0 c$) at $t_0=59.45T_0$ for the TC (red) and $t_0^{'}=58.5T_0$ for OC (black) case. Note higher $n_e$, larger $\gamma_x$ and $E_y$  resulting in more CSE. }
\label{fig:fourth}
\end{figure}

However, when laser is sufficiently strong to overcome the electrostatic field of the ions, the electrons are pushed beyond (blown out)  the foil ions, as shown in figure \ref{fig:third}(b,d). In this case the magnitude of the peak accelerating field is significantly larger, $f_C$ behind the electron bunch becomes approximately constant, so that ${f_x}|_{x_1}<{f_x}|_{x_2}$  . When $f_L$ decreases, the bunch is then accelerated back towards the laser by $f_C$, with the force distribution resulting in $a_{x_1}<a_{x_2}$ and the rear electrons catching up with the front electrons resulting  in strong bunch compression during the acceleration phase ($n_e$ increases from a maximum value  25$n_c$ before to 638$n_c$ in the blowout cycle for the TC case).  \\
The distinction between the blow-out cycle and the preceding cycles is stronger in the TC case as the reduced electron loss leads to a more favourable ion density distribution compared to rapidly decompressing OC foil. Consequently, as shown in figure \ref{fig:fourth}(a) and (b),  the peak electron density is six times larger in the TC case, due to  $0.45N_0/0.14N_0\approx 3$ times larger electron number available, and $2\times$  better compression  in the TC. Clearly denser bunches  will  result in the enhancement of coherence and intensity of the emitted attosecond pulse. Additionally, the higher peak field strength in the TC case increases the transverse acceleration and   gamma factor $\gamma_x$ (6.7 for TC and 4.7 for OC) of the bunch surface moving towards the observer (Fig. \ref{fig:fourth}(b)).  In summary, the TC case results in a substantially brighter attosecond emission due to a combination of higher electron density, larger longitudinal $\gamma_x$ and a larger transverse acceleration $E_y$. 

A simple estimate shows that this explains the bright TC emission: The coherent radiation emitted by a 1D electron layer can be written as \cite{transparency2, Bulanov2013}

 \begin{equation}
 E_{y}^{(r)}(x,t) = \frac{\pi N}{2n_{c}\lambda}{[\frac{\beta_y(t')}{1-|\beta_x(t')|}]}|_{t'=t-|x-x_e(t')|/c}. 
 \label{eq:second}
 \end{equation}
 
 At the emission time $t'=t^{'}_0$, we can simplify to 
 \begin{equation}
E_{y}^{(r)} \approx \frac{\pi N{\gamma_x}}{n_c\lambda}{\frac{a_{t^{'}_0}(t'-t^{'}_0)}{\sqrt{1+{(a_{t^{'}_0}(t'-t^{'}_0))}^2}}}|_{t'=t-|x-x'|/c}. 
\label{eq:third}
\end{equation}
 
Here we use  $\gamma_x(t')=1/\sqrt{1-\beta_x^2}\gg1$, $\beta_y(t^{'}_0)\approx0$. $\beta_x$, $p_y$ can be respectively simplified as $|\beta_x(t')|=\sqrt{1-\frac{1}{\gamma_x^2}}\approx 1-\frac{1}{2\gamma_x^2}$ and $p_y(t')\approx {\frac{dp_y}{dt'}}|_{t'=t'_{0}}(t'-t^{'}_0)$. 
\\Thus, $\beta_y(t')=p_y/\gamma \approx \frac{a_{t^{'}_0}(t'-t^{'}_0)}{\gamma_x\sqrt{1+{(a_{t^{'}_0}(t'-t^{'}_0))}^2}}$, where
\\ $a_{t^{'}_0}={\frac{dp_y}{dt'}}|_{t'=t'_{0}}$ and  $\gamma_x=\gamma/ \sqrt{(1+p_y^2)}$ are transverse acceleration and Lorentz-factor respectively.

As $a_{t^{'}_0}(t-t^{'}_0)\ll 1$, the maximum radiation intensity can be estimated by $I_{y,m}^{(r)}\approx ({\frac{\pi }{2n_{c}\lambda}})^2{(N\gamma_x a_{t^{'}_0})}^2$. I.e. the intensity will increase with the square of (1) electron number $N$, (2) longitudinal $\gamma_x$ and (3) transverse acceleration $ a_{t^{'}_0}$. Inserting the ratio of the electron number $0.45N_0/0.14N_0\approx 3.2$, averaged $\gamma_x$ of $6.7$ vs $4.7$ and the $29/22$ ratio in the transverse accelerating field $E_y$ into above equations predicts a maximum intensity $ 37\times$ higher for the TC compared to OC, consistent with the simulation results in Fig. \ref{fig:first}(b) and (c). Furthermore, the radiation pulse duration also reduces as $\gamma_x$ increases, since  $t'-t^{'}_0=dt'=dt/(1-|\beta_x(t')|)\approx 2{\gamma_x}^2 dt$ for the retarded time $t'=t-|x-x'(t')|/c$ and  ($I^{r}_y \propto |\gamma_x dt^{'}|^2$) the pulse duration scales as $ 1/{\gamma_x}^3$ resulting in a reduction from 50as to 30as. 

  \begin{figure}
 	\includegraphics[width=8cm]{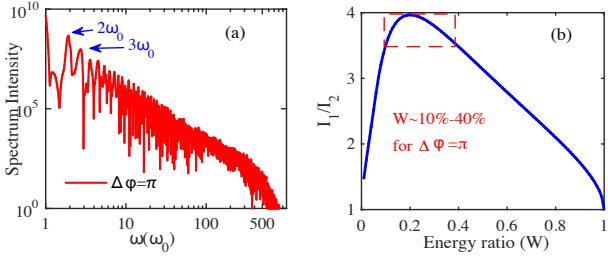}
 	\caption{(color online) (a)  Spectrum of a single attosecond pulse for the two-color case for $\Delta\phi=\pi$, $W=0.2$ described in the article. (b) Dependence of the ratio of two adjacent half-cycles  $I_1/I_2$ on the $2 \omega$ energy fraction  for $\Delta\phi=\pi$. }\label{fig:firsts}
 \end{figure}

Figure \ref{fig:firsts}(a) shows the spectrum (derived from the Fourier-transform of the reflected field) of the single attosecond pulse and Fig. \ref{fig:firsts} (b) plots the variation of  $I_1/I_2$ with $W$ for the TC case of $\Delta \phi = \pi$. Becuase the production of a single, dominant attosecond pulse from  a TC laser nano-plasma interaction depends on the temporal intensity variation of the laser, which, in turn,  depends on the relative phase $\Delta\phi$ and the energy ratio $W=E_{2\omega}/E_{\omega}$. Acording to the results in Fig. \ref{fig:first}, bright isolated attosecond pulses are produced for $\Delta\phi \approx \pi$  if the intensity ratio of two adjacent half-cycles $I_1/I_2\geq3.5$. It implys that a broad range of $W\approx 0.1-0.4$ is suitable.

\begin{figure}
	\includegraphics[width=8cm]{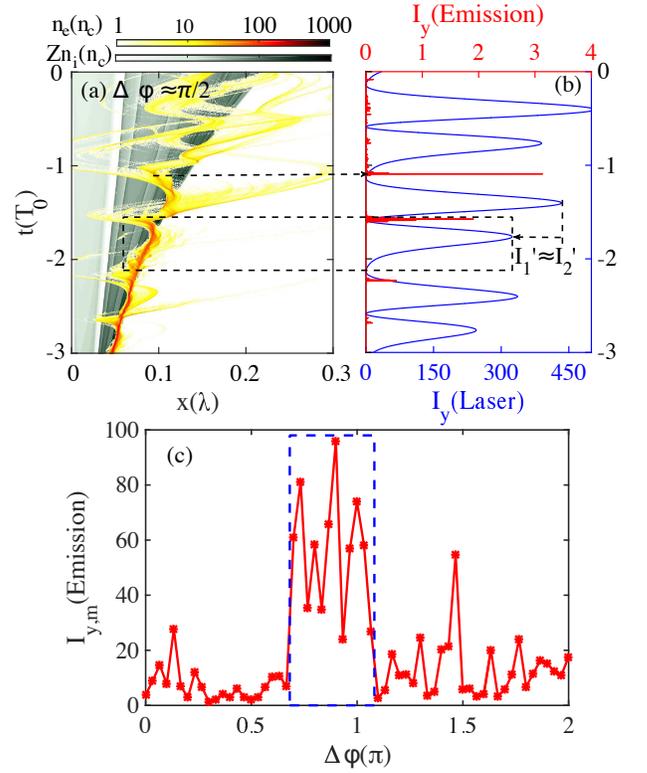}
	\caption{(color online)Phase dependence of TC interactions.  (a) Electron ($n_e$) and  ion ($n_i$) density evolution in units of $n_c$  and  (b) Intensity $I_y=|eE_y/m_e\omega_0c|^2$ of the incident laser and reflected pulse filtered from $30-200\omega_0$ both for $\Delta\phi=\pi/2$. Note that the behaviour is very similar to the OC case with a less pronounced blow out cycle and much lower peak brightness c) shows that strong pulses are obtained in a broad range around $\Delta\phi=\pi$}\label{fig:seconds}
\end{figure}

As this is a 1D simulation, this ratio applies for equal spot-size. For diffraction limited focusing to spots with size ratio $w_1=2 w_2$ the required $2\omega$ energy would be $W\approx 0.025-0.1$.  The results for two-colour simulations with non-optimal phase $\Delta\phi=\pi/2$ are shown in Figure \ref{fig:seconds} for comparison  (all other parameters are the same as in in the main text). Fig. \ref{fig:seconds}(a) corresponds to the density evolution of $n_e$ and $n_i$ and Fig. \ref{fig:seconds}(b) are the laser intensity and XUV attosecond bursts filtered also from $30-200\omega_0$. From \ref{fig:seconds}(b), it is clear that the laser field more closely resembles that of a single colour laser field, in that intensities between adjacent half-cycles are almost the same. Thus, as in the single colour case, strong electron loss occurs in every half-cycle up to the blow-out cycle and lower bunch densities, weaker compression and lower acceleration is observed compared to the optimal TC case with phase $\Delta\phi=\pi$. The XUV emission consists of an attosecond train with several times lower intensity, similar to the OC case. The phase dependence of the maximum emission intensity is shown in Fig. \ref{fig:seconds}(c). The region of strong, isolated attosecond pulse emission can be seen to be quite broad covering  $\Delta\phi\approx 0.7\pi-1.1\pi$. This insensitivity to the parameters $W$ and $\Delta \phi$  makes the observed effects experimentally easily accessible.

\section{DISCUSSION ON ROBUSTNESS TO PARAMETER VARIATION}

\begin{figure}
	\includegraphics[width=8cm]{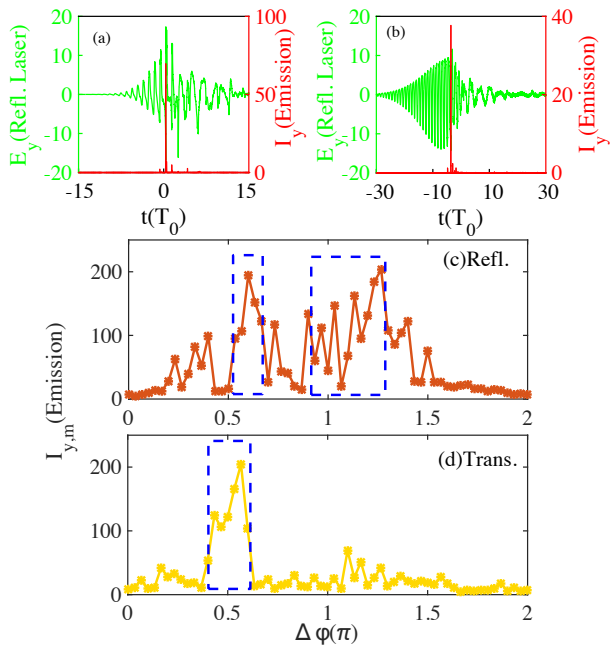}
	\caption{(color online) Parameter variations highlighting the robustness of the results. Attosecond XUV pulses overlapped with the reflected fields  for (a) an obliquely incident case ($\theta=45^{\circ}$) and (b) a longer pulse  ($\tau=50\rm{fs}$). Phase dependence of maximum emission intensity $I_{y,m}$ (also filtered from $30-200\omega_0$) for  different oblique TC cases. Bright, isolated attosecond pulses are obtained for $\Delta\phi\approx 0.8-1.1\pi$, highlighted blue-dashed box.  (c,d)  }\label{fig:thirds}
\end{figure}

This process is robust with 2D, oblique incidence and many cycle lasers as well. Figure \ref{fig:thirds}(a) shows that intense isolated attosecond pulses are also generated for oblique incidence in 1D simulations. In \ref{fig:thirds}(b) we show the target reflectivity and intensity of the attosecond pulse for a pulse duration of $\tau=50\rm{fs}$. The emission of the intense attosecond pulse is broadly unaffected by the longer pulse duration, however, as discussed previously each laser cycle results in the loss of electrons from the target and hence the resulting lower bunch density results in a slightly reduced intensity of the single attosecond pulse. 

  \begin{figure}
	\includegraphics[width=8cm]{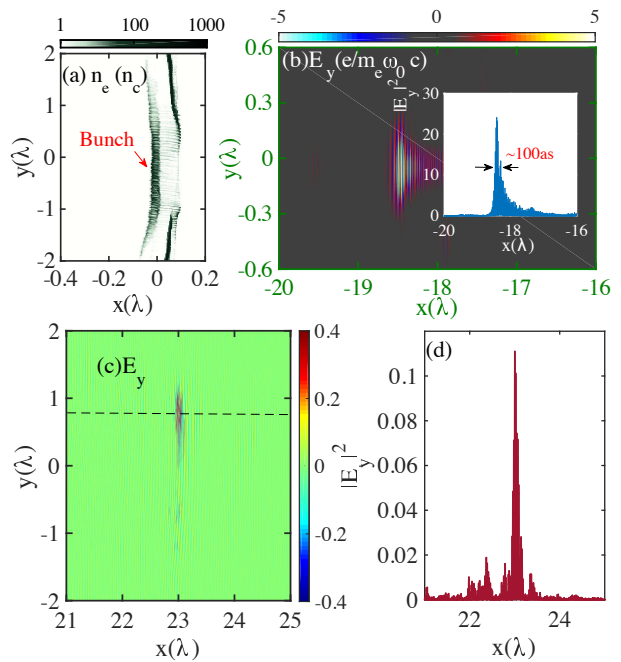}
	\caption{(color online) 2D simulations with normal incidence  with pre-plasma (scale-length $L_p=0.01\lambda$). Density distribution $n_e$ at the emission time $t=44.5T_0$, showing the formation of a dense electron bunch (top left) and the distribution and lineout of the reflected field $E_y$ filtered from $20-200\omega_0$ (top right). A substantially weaker isolated attosecond pulses is also observed in the transmitted direction (bottom left)  and intensity lineout along the dashed-line (bottom right).}\label{fig:fourths}
\end{figure}

2D PIC simulations with a pre-plasma scale-length of $L_p=0.01\lambda$ ($n_0\exp(-x/L_p)$)) were also carried out to check for multi-dimensional effects. To keep the  parameter $\epsilon$ constant, the foil density and thickness were set to $n_{e,0}=300n_c$ and $d=4\rm{nm}$. The laser pulse had a Gaussian transverse profile with FWHM radius $r_0=3\lambda$. The simulation box was composed of $36\lambda\times 16\lambda$, with a reduced spatial resolution as $\lambda/500\times \lambda/500$ and only 100 particles per cell. All other parameters are the same as the 1D simulations above. Figure \ref{fig:fourths}(a) shows the electron density close to the attosecond pulse emission time, with the dense electron bunch clearly visible. The emitted electric field $E_y$, filtered from $20$ to $200\omega_0$ is shown in Fig. \ref{fig:fourths} (b). As shown in the inset, 2D effects lead to a small reduction in the peak intensity, but the overall dynamics remain the same. Our 2D simulation results in Fig. \ref{fig:fourths} (c) and (d) show that a single attosecond pulse is also emitted in transmitted direction when the relative phase satisfies $\Delta\phi\approx\pi$, but the maximum is about two magnitude orders lower for the normal incidence case. 

\begin{figure}
	\includegraphics[width=8cm]{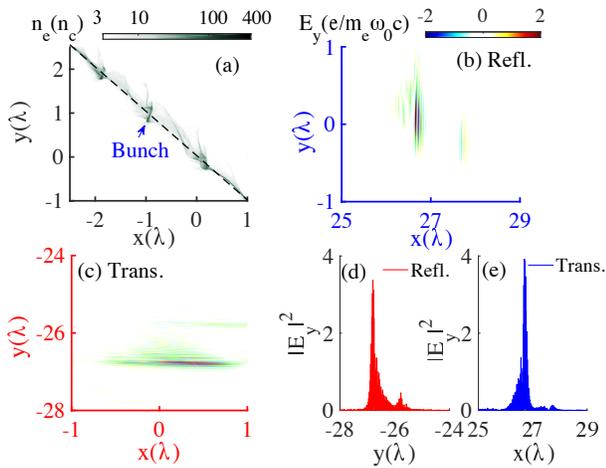}
	\caption{(color online) 2D result for $\theta=45^{\circ}$ incidence angle  with $\Delta\phi=0.7\pi$. (a) Electron density distribution at emission time showing the formation of a nanobunch. The dashed line plots the position of the initial foil. Single, giant attosecond pulse field $E_y(e/m_e\omega_0c)$ respectively in the (c) reflected and (d) transmitted direction after filtering from $20-200\omega_0$. (e)-(f) Lineout of attosecond pulse intensity in reflection and transmission.  }\label{fig:fifths}
\end{figure}

For the 2D case in oblique incidence the asymmetry observed in normal incidence is broken and for $\Delta\phi\approx 0.7\pi$  and $\theta=45^{\circ}$ equal strength attosecond bursts are emitted in both transmission and reflection (Fig. \ref{fig:fifths}(a)-(e)).

One and two dimensional (1D, 2D) simulations for oblique TC cases that  $\theta=30^{\circ}, 45^{\circ}$, and $60^{\circ}$ were also performed. To keep the blow-out cycle at the peak laser intensity, the foil density was chosen to be $400n_c$ for the oblique simulation cases, while all other parameters were kept the same as in the normal incidence cases. Note that, as the  force driving the oscillation of the electron nanobunch normal to the target is also influenced by a component of $E_y$ ($f_L\sim [E_y+v\times B_z]_{\perp}$ instead of only $v\times B_z$ for normal incidence), the optimal relative phase condition is different from  the normal incidence cases. The 1D results for $\theta=60^{\circ}$ are plotted in Fig. \ref{fig:thirds} (c) and (d) as an example. It shows that the regions of $\Delta\phi\approx\sim 0.5-0.75\pi$ or $0.85-1.3\pi$  are optimal for bright, isolated attosecond pulse in the reflected direction, and $\Delta\phi\approx 0.35-0.75\pi$ is optimal in transmission. Thus, $\Delta\phi\approx 0.5-0.7\pi$ works well in both directions. We note that while the results in 1D and 2D simulations are broadly consistent for normal incidence, the 2D oblique case shows notably weaker peak attosecond intensities (Fig. \ref{fig:fourths}). This appears to be due to 2D effects lowering the peak bunch density. This is understandable as the bunches in oblique incidence are both transversely and longitudinally narrow, making them more susceptible to deleterious effects. As shown in figure 4, the reflected radiation is well collimated with the harmonic beam growing negligibly in size over the propagation distance of 18 wavelengths. In general, as with all harmonic sources, refocusing the XUV beam requires low divergence which in turn can be controlled by using longer focal length parabolas to focus the laser - at the cost of requiring a higher laser power to achieve blow-out intensity.

\section{Summary}

In conclusion, the two-colour dynamics of laser/thin-foil interactions close to blow-out regime deviate substantially from the single frequency case. The two-colour field stabilises the target up to the blow-out point followed by a single cycle blow-out transition to transparency. The resulting strong acceleration of a dense electron bunch leads to single, dominant attosecond emission bursts due to coherent synchrotron emission, which is about 40 times stronger than those predicted for a one-color laser. This approach is robust and easly achievable for current experimental conditions. And such an intense and isolated attosecond pulse represents a significant advance in the development of attosecond sources.


\begin{thebibliography}{99}

\bibitem{Pertot:2017}Y. S. Pertot, {\it et al.}, Science, {\bf 355}(6322), 264-267 (2017).
\bibitem{Atta:2017}A. R. Atta, {\it et al.}, Science, {\bf 356}(6333), 54-59 (2017); S. X. Hu  {\it et al.} Phys. Rev. Lett., {\bf96}(7), 073004 (2006).
\bibitem{Hassner}R. H\"assner, {\it et al.}, Opt. Lett. {\bf22}, 1491–1493 (1997). 
\bibitem{Dobosz}G. Dobosz, {\it et al.}, Phys. Rev. Lett. {\bf95}, 025001 (2005).
\bibitem{Lara-Astiaso:2018}M. Lara-Astiaso, {\it et al.}, J. Phys. Chem. Lett., {\bf 9}, 4570-4577 (2018).
\bibitem{Calegari:2014}F. Cal\'egari, {\it et al.}, Science, {\bf 346}, 336-339 (2014).
\bibitem{Takahashi:2015}J. E. Takahashi, {\it et al.}, IEEE J. Sel. Top. Quantum Electron. 21, 8800112 (2015).
\bibitem{Review}F. Krausz, Phys. Scr. {\bf 91}, 063011 (2016).
\bibitem{Review1}M. Chini, {\it et al.}, Nat. Photon.{\bf 8}, 178 (2014). 
\bibitem{Review2}G. Sansone, {\it et al.}, Nat. Photon. {\bf 5}, 655 (2011).
\bibitem{application}F. Krausz, M. Ivanov, Rev. Mod. Phys. {\bf 81}, 163 (2009).
\bibitem{Isolated application} M. Chini, {\it et al.}, Phys. Rev. Lett., {\bf109}(7), 073601 (2012).
\bibitem{Tsakiris} George D Tsakiris {\it et al.},  New J. Phys. {\bf}8 19 (2006).
\bibitem{Quere}  F. Quere et al., Phys. Rev. Lett. {\bf 96}, 125004 (2006).
\bibitem{Lichters:1996} R. Lichters, {\it et al.}, Phys. Plasmas, {\bf3}, 3425-3437 (1996).
\bibitem{ROM Baeva:2006} T. Baeva, {\it et al.}, Phys. Rev. E {\bf 74}, 046404  (2006).
\bibitem{CSE Dromey} B. Dromey {\it et al.}, Nat. Phys. {\bf 8}, 804 (2012); B. Dromey, {\it et al.}, New J. Phys. {\bf15}, 015025 (2013).
\bibitem{CSE Pukhov} A. Pukhov, D. An der Br\"ugge, Plasma Phys. Control. Fusion {\bf 52}, 124039 (2010); D. an der Br\"ugge and A. Pukhov, Phys. Plasmas \textbf{17}, 033110 (2010).	
\bibitem{Cousens} S. Cousens, {\it et al.}, Phys. Rev. Lett., {\bf 116}, 083901 (2016).
\bibitem{AS train} P. M. Paul, {\it et al.}, Science, {\bf292}, 1689-1692 (2001);  P. Antoine,  {\it et al.}, Phys. Rev. Lett., {\bf 77}, 1234 (1996). 
\bibitem{Silva:2015}F. Silva, {\it et al.}, Nat. comm., {\bf 6}, 6611 (2015).
\bibitem{Intensity gating} M. Nisoli, {\it et al.}, Phys. Rev. Lett. {\bf91}, 213905 (2003); A. Jullien, et al., Appl. Phys. B {\bf93}, 433–442 (2008); J. M. Abel, {\it et al.}, Chem. Phys. {\bf366}, 9–14 (2009).
\bibitem{Polarisation gating} G. Sansone, {\it et al.}, Phys. Rev. A {\bf80}, 063837 (2009);  I. J. Sola, {\it et al.}, Nature Phys. {\bf2}, 319–322 (2006); 
\bibitem{Two color gas} J. Mauritsson, {\it et al.}, Phys. Rev. Lett., {\bf97}, 013001 (2006). T. Pfeifer, {\it et al.}, Optics Lett., {\bf31}, 975-977 (2006).
\bibitem{Gibbon:1996} P. Gibbon, Phys. Rev. Lett., {\bf76}, 50 (1996).
\bibitem{ROM Dromey} B. Dromey {\it et al.}, Nat. Phys. {\bf 2}, 456 (2006); Phys. Rev. Lett. {\bf 99}, 085001 (2007); Nat. Phys. {\bf 5}, 146 (2009).
\bibitem{Nomura:2009}Y. Nomura, {\it et al.},  Nat. Phys., {\bf5}, 124 (2009).
\bibitem{CEP}P. Heissler {\it et al.}, Phys. Rev. Lett. {\bf 108}, 235003 (2012).
\bibitem{Zhang:2018}X. Y. Zhang, {\it et al.}, Phys. Plasmas, {\bf25}, 023302 (2018).
\bibitem{Vincent:2012}H. Vincenti, {\it et al.}, Phys. Rev. Lett., {\bf108}, 113904 (2012).
\bibitem{Wheeler:2012}J. Wheeler,  {\it et al.}, Nat. Photon., {\bf 6}, 829-833, (2012).
\bibitem{Yeung:2014}M. Yeung, {\it et al.},  Phys. Rev. Lett. {\bf 112}, 123902 (2014).
\bibitem{Sergey:2008}S. Rykovannov, {\it et al.}, New J. Phys. {\bf 10}, 025025 (2008)
\bibitem{Laser system} F. Tavella, {\it et al.}, Optics Lett., {\bf32}, 2227-2229  (2007). 
\bibitem{Mirzanejad:2013} S. Mirzanejad, M. Salehi, Phys. Rev. A, {\bf 87}, 063815 (2013).
\bibitem{Edwards:2014}R. M. Edwards, {\it et al.}, Opt. Lett. {\bf 39}, 6823-6826 (2014).
\bibitem{Yeung:2017} M. Yeung, {\it et al.}, Nat. Photon., {\bf11}, 32 (2017).  
\bibitem{transparency} S. Palaniyappan, {\it et al.}, Nat. Phys., {\bf8}, 763 (2012).
\bibitem{transparency2}A. V. Vshivkov, {\it et al.}, Phys. Plasmas, {\bf5}, 2727-2741 (1998).
\bibitem{Bulanov2013} SV Bulanov   {\it et al.}, Phys. Plasmas, {\bf12} 123114 (2013)
\bibitem{Robinson} A.P.L. Robinson et al., New J. of Phys. {\bf 10}  013021(2008).
\end{thebibliography}
\end{document}